\documentstyle[12pt]{article}
\bibliography{plain}
\title{\bf Composition of Chaotic Maps with an Invariant Measure}

\vspace{20mm}

\author{M. A.
Jafarizadeh$^{a,b,c}$\thanks{E-mail:jafarzadeh@ark.tabrizu.ac.ir}
, S.Behnia$^{d,e,b}$,
 S.Khorram$^{f,b}$\thanks{E-mail:skhorram@ark.tabrizu.ac.ir} and
  H.Naghshara$^{a,b}$ \thanks{E-mail:naghshara@ark.tabrizu.ac.ir}.\\
$^a${\small Department of Theoretical Physics and Astrophysics,
Tabriz University, Tabriz 51664, Iran.} \\ $^b${\small Institute
for Studies in Theoretical Physics and Mathematics, Teheran
19395-1795, Iran.} \\ $^c${\small Pure and Applied Science
Research Center, Tabriz 51664, Iran.} \\ $^d${\small Plasma
Physics Research Center, IAU, Teheran 14835-159, Iran.} \\ $^e $
{\small Department of Physics, IAU, Urmia, Iran.} \\ $^f$ {\small
Center for Applied Physics Research, Tabriz University, Tabriz
51664, Iran.}} \pagebreak
\begin{document}
\maketitle
\vspace{15mm}
\newpage
\begin{abstract}
We generate new hierarchy of many-parameter family of maps of the
interval $[0,1]$ with an invariant measure, by composition of the
chaotic maps of reference \cite{hiera}. Using the measure, we
calculate Kolmogorov-Sinai entropy, or equivalently Lyapunov
characteristic exponent, of these maps analytically, where the
results thus obtained have been approved with numerical
simulation. In contrary to the usual one-dimensional maps and
similar to the maps of reference \cite{hiera}, these maps do not
possess period doubling or period-n-tupling cascade bifurcation to
chaos, but they have single fixed point attractor at certain
region of parameters values, where they bifurcate directly to
chaos without having period-n-tupling scenario exactly at these
values of parameter whose Lyapunov characteristic exponent begins
to be positive.\\\\ {\bf Keywords:Chaos, Invariant measure,
Entropy, Lyapunov characteristic exponent, Ergodic dynamical
systems }.\\
 {\bf PACs
numbers:05.45.Ra, 05.45.Jn, 05.45.Tp }
\end{abstract}
\pagebreak \vspace{7cm}
\section{INTRODUCTION}
In the past twenty years dynamical systems, particularly one
dimensional iterative maps have attracted much attention and have
become an important area of research activity. One of the
landmarks in it was introduction of the concept of
Sinai-Ruelle-Bowen (SRB) measure or natural invariant
measure\cite{sinai, jakob}. This is, roughly speaking, a measure
that is supported on an attractor and also describe the statistics
of the long time behavior of the orbits for almost every initial
condition in the corresponding basin of attractor . This measure
can be obtained by computing the fixed density of the so called
Frobenius-Perron operator which can be viewed as a
differential-integral operator, hence, exact determination of
invariant measure of dynamical systems is rather a nontrivial
task, such that invariant measure of few dynamical systems such as
one-parameter family one-dimensional piecewise linear maps
\cite{tasaki,tasaki1} including Baker and tent maps or unimodal
maps such as logistic map for certain values of its parameter, can
be derived analytically. In most  cases only numerical algorithms,
as an example Ulam's method\cite{froy,froy1,blank} are used for
computation of fixed densities of Frobenius-Perron operator.\\
Authors in reference \cite {hiera} have given hierarchy of one
parameter family of nonlinear maps of interval $[0,1]$ with an
invariant measure. Here in this paper we generate new hierarchy of
many-parameter family of maps of the interval $[0,1]$ with an
invariant measure, by composition of the chaotic maps of reference
\cite{hiera}. These maps are also  defined as ratio of
polynomials, where we have derived analytically their invariant
measure for  arbitrary values of the parameters. Using this
measure, we have calculated analytically, Kolmogorov-Sinai entropy
or equivalently positive Lyapunov characteristic exponent of these
maps, where the numerical simulation approve the analytic
calculation. Also it is shown that just like the  maps of
reference \cite{hiera}, they possess very peculiar property, that
is, contrary to the usual, these maps do not possess period
doubling or period-n-tupling cascade bifurcation to chaos, but
instead they have single fixed point attractor at certain region
of parameters values, where they bifurcate directly to chaos
without having period-n-tupling scenario exactly at these values
of parameter whose Lyapunov characteristic exponent begins to be
positive.\\ The paper is organized as follows: In section II we
introduce new hierarchy of family of many-parameters maps by
composition of the chaotic maps of refernce \cite{hiera}. In
Section III we show that the proposed anzats for the invariant
measure of these maps are eigenfuntion of Ferobenios-Perron
operator with largest eigenvalue $1$. Then in section IV using
this measure we calculate Kolmogorov-Sinai entropy of these maps
for an arbitrary value of parameters. In section V we compare
analytic calculation with the numerical simulation. Paper ends
with a brief conclusion.
\section{Many-Parameter Families of Chaotic Maps }
Let us consider the one-parameter families of chaotic maps of the
interval $[0,1]$ given  in reference \cite{hiera}, defined as the
ratio of polynomials of degree $N$: $$
\Phi_{N}(x,\alpha)=\frac{\alpha^2\left(1+(-1)^N{
}_2F_1(-N,N,\frac{1}{2},x)\right)}
{(\alpha^2+1)+(\alpha^2-1)(-1)^N{ }_2F_1(-N,N,\frac{1}{2},x)}$$
 \begin{equation}
  =\frac{\alpha^2(T_N(\sqrt{x}))^{2}}{1+(\alpha^2-1)(T_N(\sqrt{x})^{2})}\quad ,
  \end{equation}
where $N$ is an integer greater than one. Also $$
_2F_1(-N,N,\frac{1}{2},x)=(-1)^{N}\cos{(2N\arccos\sqrt{x})}=(-1)^{N}T_{2N}(\sqrt{x})
$$ is hypergeometric polynomials of degree $N$ and
$T_{N}(x)(U_{N}(x))$ are Chebyshev polynomials of type I (type
II)\cite{wang}, respectively. Obviously these map the unit
interval $[0,1]$ into itself.
 $ \Phi_{N}(x,\alpha)$ is (N-1)-model map, that is it has
 $(N-1)$ critical points in unit interval $[0,1]$,(see Figure 4) since its
 derivative is proportional to derivative of hypergeometric
 polynomial $_2F_1(-N,N,\frac{1}{2},x)$ which is itself a hypergeometric
 polynomial of degree $(N-1)$, hence it has
 $(N-1)$ real roots in unit interval $[0,1]$. Defining Shwarzian
 derivative\cite{dev} ${S\Phi_N(x,\alpha)}$ as: $$
S\left(\Phi_N(x,\alpha)\right)=\frac{\Phi_{N}^{\prime\prime\prime}
(x,\alpha)}{\Phi_{N}^{\prime}(x,\alpha)}-\frac{3}{2}\left(\frac{\Phi_{N}^{\prime\prime}(x,\alpha)}{\Phi_{N}^{\prime}(x,\alpha)}\right)^2=\left(\frac{\Phi_{N}^{\prime\prime}(x,\alpha)}{\Phi_{N}^{\prime}(x,\alpha)}\right)^{\prime}-\frac{1}{2}
\left(\frac{\Phi_{N}^{\prime\prime}(x,\alpha)}{\Phi_{N}^{\prime}(x,\alpha)}\right)^2,
$$
 with a prime denoting a single differentiation with respect to variable $x$, one can show that\cite{hiera}:
$$
S\left(\Phi_{N}(x,\alpha)\right)=S\left(_2F_1(-N,N,\frac{1}{2},x)\right)\leq0.
$$
 Therefore, the maps $\Phi_{N}(x)$ have at most $N+1$
attracting periodic orbits\cite{dev}. As it is shown in reference
\cite{hiera}, these maps have only  single period one stable fixed
points.\\ Using the above hierarchy of family one-parameter of
maps we can generate new hierarchy of family many-parameters
chaotic maps with an invariant measure simply from the composition
of these maps. Hence considering the functions
$\Phi_{N_{k}}(x,\alpha_k),k=1,2,\cdots,n$ we denote their
composition by:
$\Phi_{N_{1},N_{2},\cdots,N_{n}}^{\alpha_1,\alpha_2,\cdots,\alpha_n}(x)$
which can be written in terms of  them in the following form:
\begin{eqnarray}
\nonumber\Phi_{N_1,N_2,\cdots,N_n}^{\alpha_1,\alpha_2,\cdots,\alpha_n}(x)=
\overbrace{\left(\Phi_{N_1}\circ\Phi_{N_2}\circ\cdots\circ\Phi_{N_n}(x)\right)}^n=
\\
\Phi_{N_1}(\Phi_{N_2}(\cdots(\Phi_{N_n}(x,\alpha_n),\alpha_{(n-1)})\cdots,\alpha_2),
\alpha_1)
\end{eqnarray}
Since these maps consist of the composition of the
$(N_{k}-1)$-modals $(k=1,2,\cdots,n)$ maps with negative Shwarzian
derivative, therefore, they are $(N_1N_2 \cdots N_{n}-1)-$ modals
map and their Shwarzian  derivative is negative too \cite{dev}.
 Therefore these  maps  have at most $ N_1N_2 \cdots N_n+1$
attracting periodic orbits\cite{dev}. As we will show below in
this section, these maps have only a single period one stable
fixed points. Since, denoting m-composition of these functions by
$ \Phi^{(m)} $, it is straightforward to show that the derivative
of $ \Phi^{(m)} $ at its possible $ m\times n $ periodic points of
an m-cycle:
$x_{\mu,k+1}=\Phi_{N_k}(x_{\mu,k},\alpha_{k}),x_{1,\mu+1}
=\Phi_{N_n}(x_{n,\mu},\alpha_N),\mu=1,2\cdots,m\quad
\mbox{and}\quad k=1,2,\cdots,n $ and $
x_{1,1}=\Phi_{N_n}(x_{m,n},\alpha_n)$ is
\begin{equation}
\mid\frac{d}{dx}\Phi^{(m)}\mid=
=\prod_{\mu=1}^{m}(\prod_{k=1}^{n}\mid\frac{N_k}{\alpha_k}(\alpha_k^{2}+(1-\alpha_k^{2})x_{\mu,k})\mid,
\end{equation}
since for $x_{\mu,k}\in [0,1]$ we have:
$$min(\alpha_k^{2}+(1-\alpha_k^{2}x_{\mu,k}))=min(1,\alpha_k^{2}),$$
therefore,
$$min\mid\frac{d}{dx}\Phi^{(m)}\mid=\prod_{k=1}^{n}\left(\frac{N_k}{\alpha_k}min(1,\alpha_k^{2})\right)^{m}.$$
Hence the above expression is definitely  greater than one for $
\prod_{k=1}^{n}\frac{1}{N_k}< \prod_{k=1}^{n}\alpha_k
<\prod_{k=1}^{n} N_k $, that is, these maps do not have any kind
of m-cycle or periodic orbits in the region of the parameters
space defined by $\prod_{k=1}^{n} \frac{1}{N_k}<\prod_{k=1}^{n}
\alpha_k <\prod_{k=1}^{n} N_k $, actually they are ergodic in this
region of the  parameters space. From (2-3) it follows that
$\mid\frac{d}{dx}\Phi^{(m)}\mid$ at $ m \times n $ periodic points
of the m-cycle belonging to interval [0,1], varies between
$\prod_{k=1}^{n}{(N_k\alpha_k)}^{m}$ and
$\prod_{k=1}^{n}{(\frac{N_k}{\alpha_k})}^{m}$ for
$\prod_{k=1}^{n}\alpha_k<\prod_{k=1}^{n}\frac{1}{N_k}$ and between
$\prod_{k=1}^{n}(\frac{N_k}{\alpha_k})^{m}$ and
$\prod_{k=1}^{n}{(N_k\alpha_k)}^{m}$ for
$\prod_{k=1}^{n}\alpha_k>\prod_{k=1}^{n}N_k$, respectively.\\ From
the definition of these maps, we see that definitely $x=1$ and
$x=0$ (in special case of odd integer values of
$N-1,N_2,\cdots,N_n$ ) belong to one of the m-cycles.\\ For
$\prod_{k=1}^{n}\alpha_k<\prod_{k=1}^{n}\frac{1}{N_k}
(\prod_{k=1}^{n}\alpha_k>\prod_{k=1}^{n}N_k)$, the formula $(2-3)$
implies that for those cases in which $x=1 (x=0)$ belongs to one
of m-cycles we will have $\mid\frac{d}{dx}\Phi^{(m)}\mid<1$, hence
the curve of $\Phi^{(m)}$ starts at $x=1 (x=0)$ beneath the
bisector and then crosses it at the previous (next) periodic point
with slope greater than one (see Fig. 1), since the formula
$(2-3)$ implies that the slope of fixed points increases with the
decreasing (increasing) of $\mid x_{\mu,k}\mid$, therefore at all
periodic points of n-cycles except for $x=1 (x=0)$ the slope is
greater than one that is they are unstable, this is possible only
if $x=1 (x=0)$ is the only period one fixed point of these maps.\\
Hence all m-cycles except for possible period one fixed points
$x=1$ and $x=0$ are unstable. \\Actually, the fixed point $ x=0 $
is the stable fixed point of these maps in the regions of the
parameters spaces defined by $\alpha_k>0,k=1,2,\cdots,n$ and
$\prod_{k=1}^{n}\alpha_k<\prod_{k=1}^{n}\frac{1}{N_k}$ only for
odd integer values of $N_1,N_2,\cdots,N_n$, however, if one of the
integers $N_k,k=1.2,\cdots,n$ happens to be even, then the $x=0$
will not  be a stable fixed point anymore. But, the fixed point $
x=1 $ is stable fixed point of these maps in the regions of the
parameters spaces defined by
$\prod_{k=1}^{n}\alpha_k>\prod_{k=1}^{n}N_k$ and
$\alpha_k<\infty,k=1,2,\cdots,n$ for all integer values of
$N_1,N_2,\cdots,N_n$.
 \\ As an example we give below some of these maps:
\begin{equation}
 \phi_{2,2}^{\alpha_{1},\alpha_{2}}(x) ={\frac {{\alpha_{{1}}}^{2}\left
(4\,x\left (x-1\right )+\left (2\,x-1 \right
)^{2}{\alpha_{{2}}}^{2}\right )^{2}}{{\alpha_{{1}}}^{2}\left
(4\,x\left (x-1\right )+\left (2\,x-1\right
)^{2}{\alpha_{{2}}}^{2} \right )^{2}+h1}}
\end{equation}
\begin{equation} \phi_{2, 3}^{\alpha_{1},\alpha_{2}}(x)
=\frac
{{\alpha_{1}}^{2}\left(\left(x-1\right)\left(4\,x-1\right)^{2}+x\left(4\,x-3\right)^{2}{\alpha_{2}}^{2}\right)^{2}}{{\alpha_{1}}^{2}\left(\left(x-1\right)\left(4\,x-1\right)^{2}+x\left(4\,x-3\right)^{2}{\alpha_{{2}}}^{2}\right)^{2}+h2}
\end{equation}
\begin{equation}
\phi_{{3,2}}^{\alpha_{1},\alpha_{2}}(x)={\frac
{{\alpha_{2}}^{2}\left( (x-1 )\left (4\,x-1\right )^{2}+x \left
(4\,x-3\right )^{2}{\alpha_{1}}^{2}\right )^{2}}{{\alpha_{{2}
}}^{2}\left (\left (x-1\right )\left (4\,x-1\right )^{2}+x\left
(4\,x -3\right )^{2}{\alpha_{{1}}}^{2}\right )^{2}+h3 }}
\end{equation}
\begin{equation}
\phi_{{3,3}}^{\alpha_{1},\alpha_{2}}(x)=\frac
{{\alpha_{1}}^{2}{\alpha_{2}}^{2}x\left (4\,x-3\right )^{2}\left
(3 \,\left (x-1\right )\left (4\,x-1\right )^{2}+x\left
(4\,x-3\right )^{2}{\alpha_{2}}^{2}\right )}{-\left (x-1\right
)^{3}\left (4\,x-1 \right )^{6}+3\,x\left
(3\,{\alpha_{1}}^{2}-2\right )\left (4\,x-3 \right )^{2}\left
(x-1\right )^{2}\left (4\,x-1\right )^{4}{\alpha_{2}}^{2}+h4}
\end{equation}
where : $$ h1=-16\,{\alpha_{{2}}}^{2} (2\,x-1 )^{2}x(x-1)$$
$$h2=-4\,x(x-1)(4\,x-1)^{2}(4\,x-3)^{2}{\alpha_{2}}^{2} $$
 $$h3=4\,x (x-1)(4\,x-1 )^{2}(4\,x-3)^{2}{\alpha_{1}}^{2}$$
 $$h4=3\,{x}^{2}\left (-3+2\,{\alpha_{1}}^{2}\right )\left (x-1
\right )\left (4\,x-1\right )^{2}\left (4\,x-3\right
)^{4}{\alpha_{2}}^{4}+{\alpha_{1}}^{2}{x}^{3}\left (4\,x-3\right
)^{6}{\alpha_{2} }^{6}$$
 Below we also introduce their
conjugate or isomorphic maps which will be very useful in
derivation of their invariant measure and calculation of their
KS-entropy in the next section. Conjugacy means that the
invertible map $ h(x)=\frac{1-x}{x} $ maps $ I=[0,1]$ into $
[0,\infty) $ and transforms  maps ${\Phi}_{N_k}(x,\alpha_k)$ into
$\tilde{\Phi}_{N_k}(x,\alpha_k)$ defined as:$$
\tilde{\Phi}_{N_k}(x,\alpha_k)=h\circ\Phi_{N_k}(x,\alpha_k)\circ
h^{(-1)}=\frac{1}{\alpha_k^{2}}\tan^{2}(N_k\arctan\sqrt{x})$$
Hence this transforms the maps
$\Phi_{N_{1},N_{2},\cdots,N_{n}}^{\alpha_1,\alpha_2,\cdots,\alpha_n}(x)$
into
$\tilde{\Phi}_{N_{1},N_{2},\cdots,N_{n}}^{\alpha_1,\alpha_2,\cdots,\alpha_n}(x)$
defined as:\\
\\
\\
\\
$$\tilde{\Phi}_{N_1,N_2,\cdots,N_n}^{\alpha_1,\alpha_2,\cdots,\alpha_n}(x)$$
$$ \frac{1}{\alpha_1^2}\tan^{2}(N_1\arctan\sqrt\circ
\frac{1}{\alpha_2^2}\tan^{2}(N_2\arctan\sqrt\circ\cdots\circ\
\frac{1}{\alpha_n^2}\tan^{2}(N_n\arctan\sqrt x)))
$$\begin{equation} \frac{1}{\alpha_1^2}\tan^{2}(N_1\arctan\sqrt{
\frac{1}{\alpha_2^2}\tan^{2}(N_2\arctan\sqrt{\cdots
\frac{1}{\alpha_n^2}\tan^{2}(N_n\arctan\sqrt{x})}\cdots)})
\end{equation}
 \vspace{5mm}

\section{INVARIANT MEASURE }
\setcounter{equation}{0}
 Dynamical systems, even apparently simple
dynamical systems as those described by maps of an interval, can
display a rich variety of different asymptotic behavior. On
measure theoretical level these types of behavior are described by
SRB \cite{sinai} or invariant measure describing statistically
stationary states of the system. The probability measure $\mu$  on
$[0,1]$ is called an SRB or invariant measure of the maps
$\Phi_{N_{1},N_{2},\cdots,N_{n}}^{\alpha_1,\alpha_2,\cdots,\alpha_n}(x)$
 given
in $(2-2)$, if it is
$\Phi_{N_{1},N_{2},\cdots,N_{n}}^{\alpha_1,\alpha_2,\cdots,\alpha_n}(x)$-invariant
and absolutely continuous with respect to Lebesgue measure. For
deterministic system such as these composed maps, the
$\Phi_{N_{1},N_{2},\cdots,N_{n}}^{\alpha_1,\alpha_2,\cdots,\alpha_n}(x)$-invariance
 means that its invariant measure $\mu_{\Phi_{N_{1},N_{2},\cdots,N_{n}}^
 {\alpha_1,\alpha_2,\cdots,\alpha_n}}(x)$ fulfills the following
formal Ferbenius-Perron integral equation
$$\mu_{\Phi_{N_{1},N_{2},\cdots,N_{n}}^{\alpha_1,\alpha_2,\cdots,\alpha_n}}(y)=\int_{0}^{1}\delta(y-\Phi_{N_{1},N_{2},\cdots,N_{n}}^{\alpha_1,\alpha_2,\cdots,\alpha_n}(x)
)\mu_{\Phi_{N_{1},N_{2},\cdots,N_{n}}^{\alpha_1,\alpha_2,\cdots,\alpha_n}}(x)dx.$$
This is equivalent to:
\begin{equation}
\mu_{\Phi_{N_{1},N_{2},\cdots,N_{n}}^{\alpha_1,\alpha_2,\cdots,\alpha_n}}(y)=\sum_{x\in\ \Phi_{N_{1},N_{2},\cdots,N_{n}}^{\alpha_1,\alpha_2,\cdots,\alpha_n}(y)
}\mu_{\Phi_{N_{1},N_{2},\cdots,N_{n}}^{\alpha_1,\alpha_2,\cdots,\alpha_n}}(x)\frac{dx}{dy}\quad,
\end{equation}
defining the action of standard Ferobenius-Perron operator for the
map $\Phi(x)$ over a function as:
\begin{equation}
P_{\Phi_{N_{1},N_{2},\cdots, N_{n}}^{\alpha_{1},\alpha_{2},\cdots,\alpha_{n}}
}f(y)=\sum_{x\in
\Phi_{N_{1},N_{2},\cdots,N_{n}}^{\alpha_{1},\alpha_{2},\cdots,\alpha_{n}}
(y)}f(x)\frac{dx}{dy}\quad.
\end{equation}
We see that, the invariant measure
$\mu_{\Phi_{N_{1},N_{2},\cdots,N_{n}}^{\alpha_1,\alpha_2,\cdots,\alpha_n}}(x)$
is given as the eigenstate of the Frobenius-Perron operator
$P_{\Phi_{N_{1},N_{2},\cdots,N_{n}}^{\alpha_1,\alpha_2,\cdots,\alpha_n}}$
corresponding to the largest eigenvalue 1.\\ As we will prove
below the measure
$\mu_{\Phi_{N_{1},N_{2},\cdots,N_{n}}^{\alpha_1,\alpha_2,\cdots,\alpha_n}
}(x,\beta)$ defined as:
\begin{equation}
\frac{1}{\pi}\frac{\sqrt{\beta}}{\sqrt{x(1-x)}(\beta+(1-\beta)x)}\quad,
\end{equation}
is the invariant measure of the maps
$\Phi_{N_{1},N_{2},\cdots,N_{n}}^{\alpha_1,\alpha_2,\cdots,\alpha_n}(x)$
 provided that the parameter $\beta$ is positive and fulfills the following relation:
\begin{equation}
\prod _{k=1}^{n}\alpha_k
\times\frac{A_{N_{n}}(\frac{1}{\beta})}{B_{N_{n}}(\frac{1}{\beta})}\times
\frac{A_{N_{n-1}}(\frac{1}{\eta_{N_{n}}^{\alpha_{n}}(\frac{1}{\beta})}}
{B_{N_{n-1}}(\frac{1}{\eta_{N_{n}}^{\alpha_{n}}(\frac{1}{\beta})})}
\times \frac{A_{N_{n-2}}(\frac{1}{\eta_{N_{n-1},N_{n}}^
{\alpha_{n-1},\alpha_{n}}(\frac{1}{\beta})})}
{B_{N_{n-2}}(\frac{1}{\eta_{N_{n-1},N_{n}}^
{\alpha_{n-1},\alpha_{n}}(\frac{1}{\beta})})} \times \cdot \times
\frac{A_{N_{1}}(\frac{1}{\eta_{N_{2},N_{3},\cdots,N_{n}}^
{\alpha_{2},\alpha_{3},\cdots,a_{n}}(\frac{1}{\beta})})}
{B_{N_{1}}(\frac{1}{\eta_{N_{2},N_{3},\cdots,N_{n}}^
{\alpha_{2},\alpha_{3},\cdots,\alpha_{n}}(\frac{1}{\beta})})} =1
\end{equation}
where the polynomials $ A_{N_{k}}(x)$ and $ B_{N_{k}}(x)$
$(k=1,2,\cdots,n)$ are defined as:
 $$
 A_{N_{k}}(x) =\sum_{l=0}^{[ \frac{N_k}{2}]}C_{2l}^{N_k}x^{l},
$$
 $$
 B_{N_{k}}(x) =\sum_{l=0}^{[ \frac{N_k-1}{2}]}C_{2l+1}^{N_k}x^{l},
$$
 where $[\quad]$ means greatest integer part. Also the functions
$\eta_{N_{n}}^{\alpha_{n}}(\frac{1}{\beta})$,
${\eta_{N_{n-1},N_{n}}^{\alpha_{n-1},\alpha_{n}}}(\frac{1}{\beta}),
\cdots$ and
${\eta_{N_{2},N_{3},\cdots,N_{n}}^{\alpha_{2},\alpha_{3},\cdots,\alpha_{n}}}(\frac{1}{\beta})$
are defined in the following form: $$
 \nonumber \eta_{N_{n}}^{\alpha_{n}}(\frac{1}{\beta})=
\beta(\frac{\alpha_nA_{N_{n}}(\frac{1}{\beta})}
{B_{N_{n}}(\frac{1}{\beta})})^2$$ $$
\nonumber\eta_{N_{n-1},N_{n}}^{\alpha_{n-1},\alpha_{n}}(\frac{1}{\beta})=
\beta(\frac{\alpha_{n-1}A_{N_{n-1}}(\frac{1}{\eta_{N_{n}}^
{\alpha_{n}}(\frac{1}{\beta})})}
{B_{N_{n-1}}(\frac{1}{\eta_{N_{n}}^{\alpha_{n}}(\frac{1}{\beta})})})^2$$
$$ \nonumber \cdots\cdots\cdots\cdots\cdots\cdots\cdots$$ $$
\nonumber \cdots\cdots\cdots\cdots\cdots\cdots\cdots$$ $$
\nonumber\eta_{N_{2},N_{3},\cdots,N_{n}}^{\alpha_{2},\alpha_{3},\cdots,\alpha_{n}}
(\frac{1}{\beta})=
\beta(\frac{\alpha_2A_{N_{2}}(\frac{1}{\eta_{N_{3},N_{4},\cdots,N_{n}}^
{\alpha_{3},\alpha_{4},\cdots,\alpha_{n}}(\frac{1}{\beta})})}
{B_{N_{2}}(\frac{1}{\eta_{N_{3},N_{4},\cdots,N_{n}}^
{\alpha_{3},\alpha_{4},\cdots,\alpha_{n}}(\frac{1}{\beta})})})^2
\nonumber$$  As we see the above measure is defined only for
$\beta > 0$ hence, from the
 relations $(3-4)$, it follows that these maps  are ergodic in the region of
 the parameter space which leads to positive solution of
$\beta$. Taking the limits of $\beta\longrightarrow 0_+$ and
$\beta\longrightarrow \infty$ in the relation (3-4),respectively,
one can show that the ergodic regions are :
$\prod_{k=1}^{n}\frac{1}{N_k}< \prod_{k=1}^{n}\alpha_k
<\prod_{k=1}^{n} N_k $ for odd integer values of
$N_1,N_2,\cdots,N_n$ and $\alpha_k>0,\mbox{for} k=1,2,\cdots,n \&
\prod_{k=1}^{n}\alpha_k <\prod_{k=1}^{n} N_k $ if one of the
integers happens to become even, respectively. Out of these
regions they have only single stable fixed points.\\ In order to
prove that measure $(3-3)$ satisfies equation $(3-1)$, with $\beta
$ given by relation $(3-4)$, it is rather convenient to consider
the conjugate map: $$
\tilde{\Phi_{N_{1},N_{2},\cdots,N_{n}}}{\alpha_1,\alpha_2,\cdots,\alpha_n}(x),
$$ with measure $\tilde{\mu}_{\tilde{\Phi}_{N_1,N_2\cdots,N_n}^
{{\alpha_1,\alpha_2\cdots,\alpha_n}}}$ denoted by
$\tilde{\mu}_{\tilde{\phi}}$ related to the measure
$\mu_{\Phi_{N_1,N_2\cdots,N_n}^{{\alpha_1,\alpha_2\cdots,\alpha_n}}}$
denoted by $\mu_{\phi}$ through the following relation:
$$\tilde{\mu_{\tilde{\Phi}}}(x)=\frac{1}{(1+x)^2}\mu_{\Phi}(\frac{1}{1+x}).$$
Denoting
$\tilde{\Phi_{N_{1},N_{2},\cdots,N_{n}}}{\alpha_1,\alpha_2,\cdots,\alpha_n}(x)$
 by $y$ and inverting it, we get : $$
x_{k_1}=\tan^2(\frac{1}{N_1}\arctan\sqrt{y\alpha_1^2}+\frac{k_1\pi}{N_1})\quad\quad
k_1=1,..,N_1.$$ $$
x_{k_1,k_2}=\tan^2(\frac{1}{N_2}\arctan\sqrt{x_{k_1}\alpha_2^2}+\frac{k_2\pi}{N_2})\quad\quad
k_2=1,..,N_2.$$ $$
\cdots\cdots\cdots\cdots\cdots\cdots\cdots\cdots\cdots$$ $$
\cdots\cdots\cdots\cdots\cdots\cdots\cdots\cdots\cdots$$ $$
x_{k_1,k_2,\cdots,k_n}=\tan^2(\frac{1}{N_n}\arctan\sqrt{x_{k_1,k_2,\cdots,k_{n-1}}\alpha_n^2}+\frac{k_n\pi}{N_n})\quad\quad
k_1=1,..,N_1. $$
 Then, taking derivative of $x_{k_1,k_2,\cdots,k_n}$ with respect to $y$, we obtain:
$$
 \nonumber \mid\frac{dx_{k1,k_2,\cdots,k_n}}{dy}\mid=
(\prod_{k=1}^{n}\frac{\alpha_k}{N_k})
\sqrt{\frac{x_{k_1,k_2,\cdots,k_n}}{y}}$$ \begin{equation}
\frac{(1+x_{k_1,k_2,\cdots,k_n})(1+x_{k_2,k_3,\cdots,k_n})
\cdots(1+x_{k_{n-1},k_n})(1+x_{k_n})}
{(1+\alpha_n^2x_{k_2,k_3,\cdots,k_n})
(1+\alpha_{n-1}^2x_{k_3,k_4,\cdots,k_n})\cdots
(1+\alpha_3^2x_{k_{n-1},k_n}) (1+\alpha_2^2x_{k_n})
(1+\alpha_1^2y)}.\end{equation} In derivation of the formula we
have used the chain rule properties of the derivative of composite
functions.\\ Substituting the above result in equation $(3-1)$, we
have: $$ \nonumber
\tilde{\mu}_{\tilde{\Phi}}(y)\sqrt{y}(1+\alpha_1^2y)=
(\prod_{k=1}^{n}\frac{\alpha_k}{N_k})\sum_{k_1}\sum_{k_2}
\cdots\sum_{k_n}\sqrt{x_{k_1,k_2,\cdots,k_n}}$$ $$ \times
\frac{(1+x_{k_1,k_2,\cdots,k_n})(1+x_{k_2,k_3,\cdots,k_n})
\cdots(1+x_{k_{n-1},k_n})(1+x_{k_n})}
 { (1+\alpha_n^2x_{k_2,k_3,\cdots,k_n})
 (1+\alpha_{n-1}^2x_{k_3,k_4,\cdots,k_n})\cdots
 (1+\alpha_3^2x_{k_{n-1},k_n})(1+\alpha_2^2x_{k_n})}
 \tilde{\mu}_{\tilde{\Phi}}(x_{k_1,k_2,\cdots,k_n}).
$$ Now,considering the following anzatz for the invariant measure
$\tilde{\mu}_{\tilde{\Phi}}(y)$:
\begin{equation}
\tilde{\mu}_{\tilde{\Phi}}(y)=\frac{1}{\sqrt{y}(1+\beta y)}\quad,
\end{equation}
then the above equation reduces to: $$
 \nonumber
\frac{1+\alpha_1^2y}{1+\beta
y}=(\prod_{k=1}^{n}\frac{\alpha_k}{N_k})$$
$$\times\sum_{k_1=1}^{N_1}
\sum_{k_2=1}^{n_2}\cdots\sum_{k_n=1}^{N_n}
\left(\frac{(1+x_{k_1,k_2,\cdots,k_n})(1+x_{k_2,k_3,\cdots,k_n})
\cdots(1+x_{k_{n-1},k_n})(1+x_{k_n})}
{(1+\alpha_n^2x_{k_2,k_3,\cdots,k_n})(1+\alpha_{n-1}^2
 x_{k_3,k_4,\cdots,k_n})\cdots(1+\alpha_3^2x_{k_{n-1},k_n})
 (1+\alpha_2^2x_{k_n})}
\right). $$
 Now, using the  formula (3-5) of
reference\cite{hiera} we obtain: $$
\frac{\alpha_n}{N_n}\sum_{k_n=1}^{N_{n}}\frac{1+x_{k_1,k_2,\cdots,k_n}}
{1+\beta x_{k_1,k_2,\cdots,k_n}} =\frac{\alpha_nA_{N_{n}}(\frac{1}
{\beta})}{B_{N_{n}}(\frac{1}{\beta})}
\frac{1+\alpha_n^{2}x_{k_1,k_2,\cdots,k_{n-1}}}
{1+\eta_{N_{n}}^{\alpha_n}(\frac{1}{\beta})
x_{k_1,k_2,\cdots,k_{n-1}}},$$ $$
 \frac{\alpha_{n-1}\alpha_{n}}{N_{n-1}N_{n}}
\sum_{k_{n-1}=1}^{N_{n-1}}
\sum_{k_n=1}^{N_{n}}\frac{(1+x_{k_1,k_2,\cdots,k_{n-1}})
(1+x_{k_1,k_2,\cdots,k_{n}})}
{(1+\alpha_{n-1}x_{k_1,k_2,\cdots,k_{n-1}})(1+\beta
x_{k_1,k_2,\cdots,k_n})}=$$ $$ \frac{\alpha_n\alpha_{n-1}
A_{N_{n}}(\frac{1}{\beta})A_{N_{n-1}}(\frac{1}
{\eta_{N_{n}}^{\alpha_n}(\frac{1}{\beta})})}
{B_{N_{n}}(\frac{1}{\beta})B_{N_{n-1}}(\frac{1}
{\eta_{N_{n}}^{\alpha_n}(\frac{1}{\beta})})}\times
\frac{1+\alpha_{n-1}^{2}x_{k_1,k_2,\cdots,k_{n-2}}} {1+
\eta_{N_{n-1},N_{n}}^{\alpha_{n-1},\alpha_n}(\frac{1}{\beta})
x_{k_1,k_2,\cdots,k_{n-2}}},$$ $$
 \cdots \cdots \cdots \cdots \cdots \cdots \cdots \cdots \cdots $$
 $$
 \cdots \cdots \cdots \cdots \cdots \cdots \cdots \cdots \cdots $$
 $$
\nonumber(\prod_{k=1}^{n}\frac{\alpha_k}{N_k})\sum_{k_1=1}^{N_1}
\sum_{k_2=1}^{n_2}\cdots$$ $$ \times\sum_{k_n=1}^{N_n}
\left(\frac{(1+x_{k_1,k_2,\cdots,k_n})(1+x_{k_2,k_3,\cdots,k_n})
\cdots(1+x_{k_{n-1},k_n})(1+x_{k_n})
 }{ (1+\alpha_n^2x_{k_2,k_3,\cdots,k_n})
 (1+\alpha_{n-1}^2x_{k_3,k_4,\cdots,k_n})
 \cdots(1+\alpha_3^2x_{k_{n-1},k_n})(1+\alpha_2^2x_k{_{n}})}
\right)=$$ $$ \prod _{k=1}^{n}\alpha_k
\times\frac{A_{N_{n}}(\frac{1}{\beta})}
{B_{N_{n}}(\frac{1}{\beta})}\times
\frac{A_{N_{n-1}}(\frac{1}{\eta_{N_{n}}^{\alpha_{n}}
(\frac{1}{\beta})})}
{B_{N_{n-1}}(\frac{1}{\eta_{N_{n}}^{\alpha_{n}}(\frac{1}{\beta})})}$$
$$ \times \frac{A_{N_{n-2}}(\frac{1}{\eta_{N_{n-1},N_{n}}^
{\alpha_{n-1},\alpha_{n}}(\frac{1}{\beta})})}
{B_{N_{n-2}}(\frac{1}{\eta_{N_{n-1},N_{n}}^
{\alpha_{n-1},\alpha_{n}}(\frac{1}{\beta})})} \times \cdot \times
\frac{A_{N_{1}}(\frac{1}{\eta_{N_{2},N_{3},\cdots,N_{n}}
^{\alpha_{2},\alpha_{3},\cdots,a_{n}}(\frac{1}{\beta})})}
{B_{N_{1}}(\frac{1}{\eta_{N_{2},N_{3},\cdots,N_{n}}
^{\alpha_{2},\alpha_{3},\cdots,\alpha_{n}}(\frac{1}{\beta})})}
\frac{1+\alpha_1^2y}{1+\eta_{N_{1},N_{2},\cdots,N_{n}}
^{\alpha_{1},\alpha_{2},\cdots,\alpha_{n}}(\frac{1}{\beta})y}. $$
Now, inserting the right hand side of last relation  in $(3-5)$,
we get: $$ \nonumber \frac{1+\alpha_1^{2}y}{1+\beta y}=\prod
_{k=1}^{n}\alpha_k
\frac{A_{N_{n}}(\frac{1}{\beta})}{B_{N_{n}}(\frac{1}{\beta})}\times
\frac{A_{N_{n-1}}(\frac{1}{\eta_{N_{n}}^{\alpha_{n}}(\frac{1}{\beta})})}
{B_{N_{n-1}}(\frac{1}{\eta_{N_{n}}^{\alpha_{n}}(\frac{1}{\beta})})}
\times$$ $$ \frac{A_{N_{n-2}}(\frac{1}{\eta_{N_{n-1},N_{n}}^
{\alpha_{n-1},\alpha_{n}}(\frac{1}{\beta})})}
{B_{N_{n-2}}(\frac{1}{\eta_{N_{n-1},N_{n}}^
{\alpha_{n-1},\alpha_{n}}(\frac{1}{\beta})})} \times \cdot \times
\frac{A_{N_{1}}(\frac{1}{\eta_{N_{2},N_{3},\cdots,N_{n}}^
{\alpha_{2},\alpha_{3},\cdots,a_{n}}(\frac{1}{\beta})})}
{B_{N_{1}}(\frac{1}{\eta_{N_{2},N_{3},\cdots,N_{n}}
^{\alpha_{2},\alpha_{3},\cdots,\alpha_{n}}(\frac{1}{\beta})})}
\frac{1+\alpha_1^2y}{1+\eta_{N_{1},N_{2},\cdots,N_{n}}^
{\alpha_{1},\alpha_{2},\cdots,\alpha_{n}}(\frac{1}{\beta})y}. $$
 We see that the above relation will hold true provided that  the parameter
 $\beta$ fulfills the relation (3-4).
\\
\\
\\
\section{KOLMOGROV-SINAI ENTROPY}
\setcounter{equation}{0} Kolomogrov-Sinai entropy (KS) or metric
entropy \cite{sinai}  measure how chaotic a dynamical system is
and it is proportional to the rate at which information about the
state of dynamical system is lost in the course of time or
iteration. Therefore, it can also be defined as the average rate
of information loss for a discrete measurable dynamical system
$(\Phi_{N_1,N_2,\cdots,N_n}^{\alpha_1,\alpha_2,\cdots,\alpha_n}(x),\mu)$,
by introducing a partition $\alpha={A_c} (n_1,.....n_{\gamma})$ of
the interval $[0,1]$ into individual laps $A_i$ one can define the
usual entropy associated with the partition by:
$$H(\mu,\gamma)=-\sum^{n(\gamma)}_{i=1}m(A_c)\ln{m(A_c)},$$ where
$m(A_c)=\int{_{n\in{A_i}}\mu(x)dx}$ is the invariant measure of $
A_i$. Defining n-th refining $\gamma(n)$ of  $ \gamma$:
$$\gamma^{n}=\bigcup^{n-1}_{k=0}(\Phi_{N_1,N_2,\cdots,N_n}^{\alpha_1,\alpha_2,\cdots,\alpha_n}(x))^{-(k)}(\gamma),$$
and defining an entropy per unit step of refining by :
$$h(\mu,\Phi_{N_1,N_2,\cdots,N_n}^{\alpha_1,\alpha_2,\cdots,\alpha_n}(x)
,\gamma)=\lim{_{n\rightarrow{\infty}}}(\frac{1}{n}H(\mu,\gamma)),$$
if the size of individual laps of $\gamma(N)$ tends to zero as n
increases, then the above entropy is known as Kolmogorov-Sinai
entropy, that is:
$$h(\mu,\Phi_{N_1,N_2,\cdots,N_n}^{\alpha_1,\alpha_2,\cdots,\alpha_n}(x))
=h(\mu,\Phi{N_1,N_2,\cdots,N_n}^{\alpha_1,\alpha_2,\cdots,\alpha_n}_(x),
\gamma).$$ KS-entropy , which is a quantitative measure of the
rate of information loss with the refining, may also be written
as:
\begin{equation}
h(\mu,\Phi_{N_1,N_2,\cdots,N_n}^{\alpha_1,\alpha_2,\cdots,\alpha_n}(x))
=\int{\mu(x)dx}\ln{\mid\frac{d}{dx}\Phi_{N_1,N_2,\cdots,N_n}^{\alpha_1,\alpha_2,\cdots,\alpha_n}
(x)\mid},
\end{equation}
which is also a statistical mechanical expression for the Lyapunov
characteristic exponent, that is, the mean divergence rate of two
nearby orbits. The measurable dynamical system
$(\Phi{N_1,N_2,\cdots,N_n}^{\alpha_1,\alpha_2,\cdots,\alpha_n}_(x),\mu)$
is chaotic for $h>0$ and predictive for $h=0$. \\ In order to
calculate the KS-entropy of the maps
$\Phi_{N_1,N_2,\cdots,N_n}^{\alpha_1,\alpha_2,\cdots,\alpha_n}(x)$,
it is rather convenient to consider their conjugate maps given by
$(2-8)$, since it can be shown that KS-entropy is a kind of
topological invariant, that is, it is preserved under conjugacy
map. Hence we have: $$
h(\mu,\Phi_{N_1,N_2,\cdots,N_n}^{\alpha_1,\alpha_2,\cdots,\alpha_n}(x))
=h(\tilde{\mu},\tilde{\Phi_{N_1,N_2,\cdots,N_n}^{\alpha_1,\alpha_2,\cdots,\alpha_n}}(x)).
$$ Using the integral $(4-1)$, the KS-entropy of
$\Phi_{N_1,N_2,\cdots,N_n}^{\alpha_1,\alpha_2,\cdots,\alpha_n}(x)$
can be written as
\\ $$
h(\mu,\Phi_{N_1,N_2,\cdots,N_n}^{\alpha_1,\alpha_2,\cdots,\alpha_n}(x))
=h(\tilde{\mu},\tilde{\Phi_{N_1,N_2,\cdots,N_n}^{\alpha_1,\alpha_2,\cdots,\alpha_n}}(x))=
$$
$$\frac{1}{\pi}\int_{0}^{\infty}\frac{\sqrt{\beta}dx}{\sqrt{x}(1+\beta
x)}\ln(\mid\frac{d}{dy_{N_2,N_3,\cdot,N_n}}(\frac{1}{\alpha_1^{2}}\tan^{2}(N_1
\arctan\sqrt{y_{N_2,N_3,\cdot,N_n}}))\times$$ $$
\frac{d}{dy_{N_3,N_4,\cdot,N_n}}(\frac{1}{\alpha_2^{2}}\tan^{2}(N_2
\arctan\sqrt{y_{N_3,N_4,\cdot,N_n}}))\cdots
\frac{d}{dx}(\frac{1}{\alpha_n^{2}}\tan^{2}(N_n
\arctan\sqrt{x}))\mid)$$\\ or
\\ $$
h(\mu,\Phi_{N_1,N_2,\cdots,N_n}^{\alpha_1,\alpha_2,\cdots,\alpha_n}(x))=
$$
$$\frac{1}{\pi}\int_{0}^{\infty}\frac{\sqrt{\beta}dx}{\sqrt{x}(1+\beta
x)}\ln(\mid\frac{d}{dy_{N_2,N_3,\cdot,N_n}}(\frac{1}{\alpha_1^{2}}\tan^{2}(N_1
\arctan\sqrt{y_{N_2,N_3,\cdot,N_n}})))+$$ $$
\frac{1}{\pi}\int_{0}^{\infty}\frac{\sqrt{\beta}dx}{\sqrt{x}(1+\beta
x)}\ln(\mid\frac{d}{dy_{N_3,N_4,\cdot,N_n}}(\frac{1}{\alpha_2^{2}}\tan^{2}(N_2
\arctan\sqrt{y_{N_3,N_4,\cdot,N_n}})))+\cdots+ $$ $$
\frac{1}{\pi}\int_{0}^{\infty}\frac{\sqrt{\beta}dx}{\sqrt{x}(1+\beta
x)}\ln(\mid\frac{d}{dy_{N_n}}(\frac{1}{\alpha_{n-1}^{2}}\tan^{2}(N_{n-1}
\arctan\sqrt{y_{N_n}})))+$$  \begin{equation}
\frac{1}{\pi}\int_{0}^{\infty}\frac{\sqrt{\beta}dx}{\sqrt{x}(1+\beta
x)}\ln(\mid\frac{d}{dx}(\frac{1}{\alpha_n^{2}}\tan^{2}(N_n
\arctan(\sqrt{x}))))\mid) \end{equation}\\ where
\begin{equation} y_{N_n}=\frac{1}{\alpha_n^{2}}\tan^{2}(N_n
\arctan(\sqrt{x}))
\end{equation} \begin{equation}
y_{N_{n-1},N_n}=\frac{1}{\alpha_{n-1}}\tan^{2}(N_{n-1}\arctan(\sqrt{y_{N_n}})))
\end{equation} $$ \cdots \cdots \cdots \cdots \cdots \cdots
\cdots \cdots \cdots$$ $$ \cdots \cdots \cdots \cdots \cdots
\cdots \cdots \cdots \cdots$$  \begin{equation}
y_{N_2,N_3,\cdot,N_n}=\frac{1}{\alpha_1^{2}}\tan^{2}(N_1
\arctan(\sqrt{y_{N_3,N_4,\cdot,N_n}})).  \end{equation}. Now, we
calculate the integrals appearing above in the expression for the
entropy, separately. The last integral in (4-2) can be calculated
with the prescription of section IV of the reference \cite{hiera}
and it reads:
\begin{equation}
\frac{1}{\pi}\int_{0}^{\infty}\frac{\sqrt{\beta}dx}{\sqrt{x}(1+\beta
x)}\ln(\mid\frac{d}{dx}(\frac{1}{\alpha_n^{2}}\tan^{2}(N_n
\arctan(\sqrt{x}))))\mid)\\
=\ln\left(\frac{N_n(1+\beta+2\sqrt{\beta})^{N_n-1}}{A_{N_{n}}
(\frac{1}{\beta})B_{N_{n}}(\frac{1}{\beta})}\right).
\end{equation}
In order to calculate the integral one before  last in (4-2), that
is:\begin{equation}
\frac{1}{\pi}\int_{0}^{\infty}\frac{\sqrt{\beta}dx}{\sqrt{x}(1+\beta
x)}\ln(\mid\frac{d}{dy_{N_n}}(\frac{1}{\alpha_{n-1}^{2}}\tan^{2}(N_{n-1}
\arctan\sqrt{y_{N_n}})))\end{equation} first we make the following
change of variable by inverting the relation(4-3):
$$x_{k_{n-1}}=\tan^2(\frac{1}{N_{n-1}}\arctan
(\sqrt{y_{N_{n}}}\alpha_{n-1}^2)+\frac{k_{n-1}
\pi}{N_{n-1}})\quad\quad k_{n-1}=1,..,N_{n-1}. $$ then the
integral (4-7) reduces to: $$
\sum_{k_{n-1}=1}^{N_{n-1}}\frac{1}{\pi}\int_{x_{k_{n-1}}^{i}}^
{x_{k_{n-1}}^{f}}\frac{\sqrt{\beta}dx_{k_{n-1}}}{\sqrt{x_{k_{n-1}}}
(1+\beta x_{k_{n-1}})}\ln(\mid \frac{d}{dy_{N_n}}
(\frac{1}{\alpha_{n-1}^{2}}\tan^{2}(N_{n-1}\arctan\sqrt{y_{N_n}}))$$
 where $x_{k_{n-1}}^{i}$ and $x_{k_{n-1}}^{f}$  ($
k_{n-1}=1,2,\cdots,N_{n-1}$) denote the initial and end points of
$k$-th branch of the inversion of function
$y_{N_n}=(\frac{1}{\alpha_n^{2}}\tan^{2}(N_n \arctan\sqrt{x}))$,
respectively. Now, inserting the derivative of $x_{k_{n-1}}$ with
respect to $ y_{{N_n}}$ in the above relation and changing the
order of sum and integration, we get: $$
 \frac{1}{\pi}\int_{0}^{\infty}\sum_{k_{n-1}=1}^{N_{n-1}}
\sqrt{\beta}dy_{N_{n}}\frac{\alpha_{n-1}\sqrt{x_{k_{n-1}}}
(1+x_{k_{n-1}})}
{N_{n-1}\sqrt{y_{n_{n}}}(1+\alpha_{n-1}^2y_{N_{n}})
{\sqrt{x_{k_{n-1}}}}(1+\beta x_{k_{n-1}})}$$
$$\times\ln(\mid\frac{d}
{dy_{N_n}}(\frac{1}{\alpha_{n-1}^{2}}\tan^{2}(N_{n-1}\arctan\sqrt{y_{N_n}}))).$$
Using the formula (3-5) of reference \cite{hiera}, it reduces to
$$ \frac{1}{\pi}\int_{0}^{\infty}
\frac{\sqrt{\beta}dy_{n}}{\sqrt{y_{n}}}
(\frac{B_{N_{n}}(\frac{1}{\beta})}{\alpha_nA_{N_{n}}(\frac{1}{\beta})}+
\beta\frac{\alpha_nA_{N_{n}}(\frac{1}{\beta})}
{B_{N_{n}}(\frac{1}{\beta})}y_{N_{n}})\ln(\mid\frac{d}{dy_{N_n}}
(\frac{1}{\alpha_{n-1}^{2}}\tan^{2}(N_{n-1}\arctan\sqrt{y_{N_n}}))).$$
Finally, calculating the above integral with the prescription of
reference\cite{hiera} we obtain:
$$\ln\left(\frac{N_{n-1}(1+\eta_{N_{n}}^{\alpha_{n}}+2\sqrt{\eta_{N_{n}}^{\alpha_{n}}})^{N_{n-1}-1}}
{A_{N_{n-1}}(\eta_{N_{n}}^{\alpha_{n}})
B_{N_{n-1}}(\eta_{N_{n}}^{\alpha_{n}})}\right).$$ Similarly, we
can calculate the other integrals appearing in the expression for
the entropy of the composed maps given in (4-2):
$$=\frac{1}{\pi}\int_{0}^{\infty}\frac{\sqrt{\beta}dx}{\sqrt{x}(1+\beta
x)}\ln(\mid\frac{d}{dy_{N_k,N_{k+1},\cdot,N_n}}(\frac{1}{\alpha_{k-1}^{2}}\tan^{2}(N_{k-1}
\arctan\sqrt{y_{N_k,N_{k+1},\cdot,N_n}}))=$$ $$
\ln\left(\frac{N_{k-1}(1+ \eta_{N_{k},N_{k-1},\cdots,N_{n}}^
{\alpha_{k},\alpha_{k+1},\cdots,\alpha_{n}}(\frac{1}{\beta})+
2\sqrt{\eta_{N_{k},N_{k-1},\cdots,N_{n}}^
{\alpha_{k},\alpha_{k+1},\cdots,\alpha_{n}}(\frac{1}{\beta})})^{N_{k-1}-1}}
{A_{N_{k-1}}(\eta_{N_{k},N_{k-1},\cdots,N_{n}}^
{\alpha_{k},\alpha_{k+1},\cdots,\alpha_{n}}(\frac{1}{\beta}))
B_{N_{k-1}}(\eta_{N_{k},N_{k-1},\cdots,N_{n}}^
{\alpha_{k},\alpha_{k+1},\cdots,\alpha_{n}})(\frac{1}{\beta})}\right).
\mbox{for} k=1,2,\cdots,n$$ Finally summing the above integral we
get the following expression for the entropy of these maps:
$$h(\mu,\Phi_{N_1,N_2,\cdots,N_n}^{\alpha_1,\alpha_2,\cdots,\alpha_n}(x))=$$
\begin{equation}
 \ln(\frac{(N_1N_2\cdots
N_n)(1+\sqrt{\beta})^{2(N_{n}-1)}
(1+\sqrt{\eta_{N_{n}}^{\alpha_{n}}(\frac{1}{\beta})})^
{2(N_{n-1}-1)}\cdots (1+\sqrt{\eta_{N_{2},N_{3},\cdots,N_{n}}^
{\alpha_{2},\alpha_{3},\cdots,\alpha_{n}}(\frac{1}{\beta})})^{2(N_{1}-1)}}
{A_{N_n}(\beta)B_{N_n}(\beta)A_{N_{n-1}}(\eta_{N_{n}}^
{\alpha_{n}}(\frac{1}{\beta}))
B_{N_{n-1}}(\eta_{N_{n}}^{\alpha_{n}}(\frac{1}{\beta}))\cdots
A_{N_{1}}(\eta_{N_{2},N_{3},\cdots,N_{n}}^
{\alpha_{2},\alpha_{3},\cdots,\alpha_{n}}(\frac{1}{\beta}))
B_{N_{1}}(\eta_{N_{2},N_{3},\cdots,N_{n}}^
{\alpha_{2},\alpha_{3},\cdots,\alpha_{n}}(\frac{1}{\beta}))}).
\end{equation}\\
Imposing the relations between the parameters
$\alpha_k,k=1,2,\cdots,n$ which are consistent with the relation
(3-4), reduces these maps to other maps of many-parameters family
of the maps with number  of the parameters less than $n$.
Particularly by imposing enough relations we can reduce them to
one-parameter family of chaotic maps with an arbitrary asymptotic
behavior as the parameter takes the limiting values. Hence we can
construct chaotic maps with arbitrary universality class. As an
illustration we consider the chaotic map
$\Phi_{2,2}^{\alpha_1,\alpha_2}(x)$. Using the formula (4-8)  we
have
$$h(\mu,\Phi_{2,2}^{\alpha_1,\alpha_2}(x)=\ln\frac{(1+\sqrt{\beta})^2
(2\sqrt{\beta}+\alpha_2(1+\beta))^2}{(1+\beta)
(4\beta+\alpha_2^2(1+\beta)^2)} $$ with the following relation
among the parameters $\alpha_1,\alpha_2$ and $\beta$: $$
\alpha_1(4\beta+\alpha_2^2(1+\beta)^2)=4\alpha_2\beta(1+\beta) $$
which is obtained from the relation(3-4). Now choosing
$\beta=\alpha_2^{\nu},\>, 0<\nu<2$, the above relation reduces to:
$$ a_1=\frac{4\alpha_2^{1+\nu}(1+\alpha_2^{\nu})}
{\alpha_2^2(1+\alpha_2^{\nu})^2+4\alpha_2^{\nu}}$$ and entropy
given by (4-8) reads: $$
h(\mu,\Phi_{2,2}^{\alpha_2}(x)=\ln\frac{(1+\alpha_2^{\frac{\nu}{2}})^2
(2\alpha_2^{\frac{\nu}{2}}+\alpha_2(1+\alpha_2^{\nu}))^2}
{(1+\alpha_2^{\nu})(4\alpha_2^{\nu}+\alpha_2^2(1+\alpha_2^{\nu})^2)}
$$ which has the following asymptotic behavior near
$\alpha_2\longrightarrow 0$ and $\alpha_2\longrightarrow\infty$:
$$ \left\{ \begin{array}{l}
h(\mu,\Phi_{2,2}^{\alpha_2}(x)\sim\alpha_2^{\frac{\nu}{2}}\quad
\quad\mbox{as}\alpha_2\longrightarrow 0 \\
h(\mu,\Phi_{2,2}^{\alpha_2}(x)\sim(\frac{1}{\alpha_2})^
{\frac{\nu}{2}}\quad\quad\mbox{as}\alpha_2\longrightarrow\infty.
\end{array}\right.$$
 The above asymptotic form indicates that, for an  arbitrary value
of $0<\nu<2$, the maps $\Phi_{2,2}^{\alpha_2}(x)$ belong to the
universality class which is different from the universality class
of chaotic maps of the reference\cite{hiera} or the universality
class of pitch fork bifurcating maps.
\\
\\
\\
 \setcounter{equation}{0} Here in this section we try to
calculate Lyapunov characteristic exponent of maps $
\Phi_{N}^{(1,2)}(x,\alpha)$, $N=1,2,....,5$ in order to
investigate these maps numerically. In fact, Lyapunov
characteristic exponent is the characteristic exponent of the rate
of average magnificent of the neighborhood of an arbitrary point
$x_{0}$  and it is denoted by $ \Lambda(x_{0}) $ which is written
as: $$
\Lambda_{N_1,N_2,\cdots,N_n}^{\alpha_1,\alpha_2,\cdots,\alpha_n}(x_{0})
=lim_{n\rightarrow\infty}\ln(\mid\overbrace{\Phi_{N_1,N_2,\cdots,N_n}^
{\alpha_1,\alpha_2,\cdots,\alpha_n}(x,\alpha) \circ
\Phi_{N_1,N_2,\cdots,N_n}^{\alpha_1,\alpha_2,\cdots,\alpha_n}
....\circ
\Phi_{N_1,N_2,\cdots,N_n}^{\alpha_1,\alpha_2,\cdots,\alpha_n}}^{n}\mid
$$
\begin{equation}
=lim_{n\rightarrow\infty}\sum_{k=0}^{n-1}\ln\mid\frac{d\Phi_{N_1,N_2,\cdots,N_n}^
{\alpha_1,\alpha_2,\cdots,\alpha_n}(x_{k},\alpha)}{dx}\mid,
\end{equation}
where $
x_{k}=\overbrace{\Phi_{N_1,N_2,\cdots,N_n}^{\alpha_1,\alpha_2,\cdots,\alpha_n}
\circ\Phi_{N_1,N_2,\cdots,N_n}^{\alpha_1,\alpha_2,\cdots,\alpha_n}
\circ....\circ\Phi_{N_1,N_2,\cdots,N_n}^{\alpha_1,\alpha_2,\cdots,\alpha_n}}
$ . It is obvious that $\Lambda^{(1,2)}(x_0)<0 $ for an attractor,
$\Lambda{N_1,N_2,\cdots,N_n}^{\alpha_1,\alpha_2,\cdots,\alpha_n}(x_{0})>0$
for a repeller and $\Lambda
{N_1,N_2,\cdots,N_n}^{\alpha_1,\alpha_2,\cdots,\alpha_n}(x_{0})=0$
for marginal situation. Also the Liapunov number is independent of
initial point $x_{0}$, provided that the motion inside the
invariant manifold is ergodic, thus
$\Lambda{N_1,N_2,\cdots,N_n}^{\alpha_1,\alpha_2,\cdots,\alpha_n}(x_{0})$
characterizes the invariant manifold of
$\Phi_{N_1,N_2,\cdots,N_n}^{\alpha_1,\alpha_2,\cdots,\alpha_n}$ as
a whole. For values of parameter $\alpha_k,k=1,2,\cdots,n$, such
that the map
$\Phi_{N_1,N_2,\cdots,N_n}^{\alpha_1,\alpha_2,\cdots,\alpha_n}$ be
measurable, Birkohf ergodic theorem implies equality of KS-entropy
and Liapunov characteristic exponent, that is:
\begin{equation}
h(\mu,\Phi_{N_1,N_2,\cdots,N_n}^{\alpha_1,\alpha_2,\cdots,\alpha_n})=
\Lambda_{N_1,N_2,\cdots,N_n}^{\alpha_1,\alpha_2,\cdots,\alpha_n}
(x_{0},\Phi_{N_1,N_2,\cdots,N_n}^{\alpha_1,\alpha_2,\cdots,\alpha_n}).
\end{equation}
Comparison of analytically calculated KS-entropy of maps
$\Phi_{N_1,N_2,\cdots,N_n}^{\alpha_1,\alpha_2,\cdots,\alpha_n}(x,\alpha)$
for $N_1=2,3$ and $N_2=2,3$, (see Figures $5,6$ and $7$ ) with the
corresponding Lyapunov characteristic exponent obtained by
simulation, indicate that in chaotic region, these maps are
ergodic as predicted by Birkohf ergodic theorem. In non chaotic
region of the parameter, Lyapunov characteristic exponent is
negative, since in this region we have only stable period one
fixed points without bifurcation. In summary, combining the
analytic discussion of section II  with the numerical simulation
we deduce that these maps are ergodic in certain region of their
parameters space as explained above and in the complementary
region of the parameters space they have only a single period one
attractive fixed point, such that in contrary to the most of usual
one-dimensional one-parameter or many-parameters family of maps
they have only a bifurcation from a period one attractive fixed
point to chaotic state or vice-versa.\\
\section{Conclusion}
\setcounter{equation}{0}
 We have given hierarchy of exactly
solvable many-parameter family of one-dimensional chaotic maps
with an invariant measure, that is measurable dynamical system
with an interesting property of being either chaotic (proper to
say ergodic ) or having  stable period one fixed point and they
bifurcate from a stable single periodic state to chaotic one and
vice-versa without having usual period doubling or
period-n-tupling scenario.
\\ Again  this interesting property is due to existence of
invariant measure for a region  of the parameters space of these
maps. Hence, to approve this conjecture, it would be interesting
to find other measurable one parameter maps, specially higher
dimensional maps, which is under investigation.

\end{document}